\documentstyle[aps,preprint]{revtex}

\begin{document}

\draft

\title{Spin-isospin textured excitations in a double layer at filling factor 
$\nu =2$}

\author{B. Paredes$^1$, C. Tejedor$^1$, L. Brey$^2$ and L. Mart\'{\i}n-Moreno$^3$}

\address{$^1$Departamento de F\'{\i}sica Te\'orica de la Materia
Condensada, Universidad Aut\'onoma de Madrid,  28049 Madrid, Spain.}

\address{$^2$Instituto de Ciencia de Materiales de Madrid (CSIC), 
Cantoblanco, 28049, Madrid, Spain.}

\address{$^3$Departamento de F\'{\i}sica de la Materia Condensada, 
ICMA(CSIC), Universidad de Zaragoza, Zaragoza 50015, Spain.}

%\date{\today}

\maketitle

\begin{abstract}

We study the charged excitations of a double layer at filling factor $2$ 
in the ferromagnetic regime. In a wide range of Zeeman and tunneling splittings
we find that the low energy charged excitations are spin-isospin textures with 
the charge mostly located in one of the layers. 
As tunneling increases, the {\it parent} spin texture in one layer becomes 
larger and it induces, in the other layer, a {\it shadow} spin texture 
antiferromagnetically coupled to the {\it parent} texture.
These new quasiparticles should be observable by measuring  
the strong dependence of its spin on tunneling and Zeeman couplings. 

\end{abstract}


%%\narrowtext

Charged excitations of a single layer quantum Hall ferromagnet at filling factor 
$\nu =1$ have been proposed to be textured quasiparticle
excitations, {\it skyrmions}\cite{sondhi}. Skyrmions are relevant 
because at $\nu$=1 the system is an itinerant ferromagnet 
in which the topological charge associated with the spin texture is 
coupled to the actual electric charge\cite{GM}.
The competition between the Zeeman splitting, $\Delta _z$, and the Coulomb 
energy determines the size and energy of the skyrmions\cite{Fertig}.
Different experimental techniques have confirmed that for typical values 
of $\Delta _z$ the skyrmions are the lowest energy charged excitations
at $\nu$=1\cite{Barret,Schmeller,Aifer,Bayot,Maude,Leadley}.
Textured charged excitations could also exist in double layer (DL) systems  
at total filling factor $\nu$=1. 
In this case the layer index plays the role of an isospin, and the 
problem can be mapped onto that of an easy plane ferromagnet. 
Then, the finite energy isospin textured charged 
excitations are {\it bimerons}\cite{bimeron}.
Alternatively, as we do here, it is possible to
map the symmetric/antisymmetric states onto two isospin orientations.

The situation becomes rather more complicated when both spin and isospin 
must be considered simultaneously. This happens, for instance, 
for a DL at $\nu =2$. This system presents a rich 
quantum phase diagram including ferromagnetic, 
canted antiferromagnetic and symmetric ground states (GS's)
\cite{Zheng,DasSarma}. 
The existence and transitions between these phases have been recently 
observed\cite{expcanted}. 
In this system the spin and the isospin are strongly mixed, and 
a question to be raised is the possible existence of spin-isospin 
textured excitations (SITE) in DL systems at $\nu$=2.
A serious difficulty in the study of charged excitations 
resides in the lack of an adequate field theory that 
could give insight on the kind of excitations to be expected. 

In this Letter, we present a variational wave-function for 
describing the charge excitations in DL systems at $\nu=2$ with 
interlayer separation $d$. In this work, we restrict 
to the regime of ferromagnetic GS. 
Our wave-function, which mixes spin and isospin textures, is built up in such  
a way that it recovers a skyrmion in one layer in the limit of zero tunneling 
splitting, $\Delta _{sas}$. We have checked the adequacy of our wave-function 
by performing a numerical diagonalization for a few electrons system.
We compute the energy, spin, isospin, and charge distribution of the charged 
excitation as a function of the DL parameters. 

Our main results are: 

{\it i)} The map of charged excitations, in terms of $\Delta _z $ and 
$\Delta _{sas}$, presents two different regions as shown in Fig. \ref{fig1}. 
In one region, labeled as the one having single-particle excitations (SPE), 
the extra charge is shared by the two layers in a single-particle 
symmetric state with spin $1/2$. In the other region, the excitation has a 
larger spin and the extra charge is mostly located in one of the layers. 
The transition between the two regions involves an abrupt change of both spin
and isospin.

{\it ii)} In the second region, labeled as the one having a SITE, the charge 
induces in one layer a {\it parent} spin texture. Its spin is larger than 
that of the skyrmion in a single layer at $\nu =1$ with the same $\Delta _z$. 
This enlargement is accompanied by the appearance of a {\it shadow}  
spin texture in the other layer. {\it Parent} and {\it shadow} spin textures 
have an in-plane antiferromagnetic coupling.
The existence of the {\it shadow} spin texture is associated with the 
existence of an isospin texture in the system so that the quasiparticle is 
a SITE.

{\it Hamiltonian.} We work in the lowest Landau level approximation where 
the single particle eigenstates in the symmetric gauge are labeled by the third 
component of the angular momentum $m$, the spin $\sigma$=$\uparrow, 
\downarrow $($\pm 1$) and the layer index $\Lambda =L,R$. 
The Hamiltonian of the DL is 
\begin{eqnarray}
&& H \! =  -\frac{\Delta_{sas}}{2} \sum_{m \sigma} \left( c^\dagger_{m \sigma L}
%%H \!& = & -\frac{\Delta_{sas}}{2} \sum_{m \sigma} \left( c^\dagger_{m \sigma L}
c_{m\sigma R}+  \mbox{h.c.}
\right)
-{{ \Delta_z} \over 2}\sum_{m\sigma \Lambda }\sigma c^\dagger_{m\sigma \Lambda }
c_{m\sigma \Lambda }
\nonumber \\
&& + \! \sum_{\{ m_j \} \sigma \sigma ' \Lambda \Lambda ' }
\frac{V^{\Lambda  \Lambda ' }_{m_1m_2m_3m_4}}{2}
c^\dagger_{m_1\sigma \Lambda  }c^\dagger_{m_2\sigma ' \Lambda '}
c_{m_3\sigma ' \Lambda '} c_{m_4\sigma \Lambda }.
\label{hamilt}
\end{eqnarray}
The interaction potential is,
$V^{\Lambda \Lambda }$=$e^2/(\varepsilon r)$  and
$V^{ \Lambda \Lambda '}$=$e^2/(\varepsilon \sqrt {r^2+d^2})$ for 
$\Lambda \! \neq \! \Lambda '$. 
Here $r=\sqrt{x^2+y^2}$ is the in-plane distance. 
Hereafter all distances will be given in units of the magnetic length
$l$ and the energies in units of $e^2/(\varepsilon l)$. 

{\it GS at $\nu=2$}. The Hartree-Fock in-plane translational invariant 
solutions of this Hamiltonian have the form,
\begin{equation}
|\Psi _{GS}\rangle  = \prod _{i=1} ^{2}
\prod _{m=0} ^{\infty}
d _{mi} ^{ \dagger} |0 \rangle \, \, \, .
\end{equation}
Here $|0 \rangle$ is the vacuum state and $d_{mi}^{ \dagger}
=\sum _{\sigma \Lambda}\alpha _{i\sigma \Lambda}c^\dagger_{m\sigma \Lambda}$ 
with the coefficients $\alpha$ determined to minimize the energy of the system. 
Note that $i$ runs from $1$ to $4$, but since we work at $\nu=2$, for each 
$m$ we only fill up the two lowest energy states. As a function of the 
parameters $d$, $\Delta _z$ and $\Delta _{sas}$, the GS given by Eq.(2)
presents ferromagnetic, canted or singlet interlayer spin 
correlations\cite{Zheng,DasSarma}. 
In this Letter, we are interested in the 
regime of parameters giving a ferromagnetic GS at $\nu$=2. 
In this case, $d_{m1}^\dagger =c_{m\uparrow L}^\dagger $, $d_{m2}^\dagger 
=c_{m\uparrow R}^\dagger $, $d_{m3}^\dagger =c_{m\downarrow L}^\dagger $ 
and $d_{m4}^\dagger =c_{m\downarrow R}^\dagger $. 

{\it Charged excitations.}
We want to write down a wave-function describing the low energy
charged excitations of the GS described by Eq.(2).
This wave-function should have the freedom to get topological charge $1$. 
The trial wave-function is allowed to get topological charge by 
mixing GS occupied orbitals with angular momentum $m$,
($d ^{\dagger} _{mj} , j=1,2$) with GS empty orbitals of angular momentum $m+1$ 
($d ^{\dagger} _{m+1j} , j=3,4$)\cite{note1}.
With this, the form of our variational trial wave-function for the 
SITE has the circularly symmetric form
\begin{eqnarray}
& & |\Psi _{qp} \rangle =
\left(\gamma _3 d_{03}^{\dagger}+\gamma _4 d_{04} ^{\dagger}\right) 
\nonumber \\ 
\times & & \prod _{i=1} ^2 \prod _{m=0} ^ {\infty} 
\left( 
\sum _{j=1}^2 \beta _{ijm} d _{mj} ^{\dagger}
+ \sum _{j=3}^4 \beta _{ijm} d _{m+1j} ^{\dagger}
\right)
| 0 \rangle 
\label{trialwf}
\end{eqnarray}
The complex parameters ${\beta}$ verify the constrains,
$\sum _{j =1} ^{4} \left ( \beta _{ijm} \right ) ^ * \beta _{i'jm} 
\! =\!  \delta _{ii'}$. The first factor in Eq.(\ref{trialwf}) stands for 
the extra electron in $|\Psi _{qp}\rangle$ with respect to $|\Psi _{GS} 
\rangle$, and therefore $|\gamma _3|^2 + |\gamma _4| ^2=1.$ 
The complex parameters $\{\gamma \}$ and $ \{ \beta \}$ are obtained 
by minimizing the energy of the system while imposing that, as $m$ increases, 
the coefficients $\beta _{i3m}$, $\beta _{i4m}$, 
$\beta _{21m}$ and $\beta _{12m}$ should decay to zero and
$\beta _{11m}$ and $\beta _{22m}$ should tend to unity in order 
to recover the shape of the GS far from the center of the excitation.

Some comments about the wave-function (\ref{trialwf}) are in order: \\ 
First, we have checked the adequacy of allowing the mixing between 
states with angular momentum $m$ and spin $\uparrow $ in each layer 
($c^{\dagger}_{m\uparrow\Lambda}$), and states with angular momentum 
$m+1$ and spin $\downarrow $ in any layer, 
($c^{\dagger}_{m+1 \downarrow \Lambda '}$). 
For this purpose, we have diagonalized numerically  
the Hamiltonian for 9 electrons in a double layer laterally confined by a 
parabolic potential. The potential has the adequate curvature to simulate a
situation similar to the one we have in the thermodynamic limit\cite{note2}. \\
Second, the wave-function (\ref{trialwf})  
takes into account that the states belong to a four dimensional space
corresponding to the spin and isospin degrees of freedom. 
This is the reason for the appearance of quartets in our trial 
wave-function for the SITE. \\ 
Third, in the limit of $d=0$ and $\Delta _z=\Delta _{sas}$ our trial 
wave-function (\ref{trialwf}) is a good candidate to describe the skyrmions 
of the $SU(4)$ model\cite{Arovas}. \\
Fourth, when the electron layers are decoupled ($\Delta _{sas}$=0), the GS  
is always ferromagnetic. If the extra electron is located 
in a single layer, (\ref{trialwf}) takes the 
form of the wave-function of a skyrmion in that layer at $\nu=1$\cite{Fertig}.

Due to the form of $| \Psi _{qp} \rangle$ the minimization of the energy 
only requires diagonalization of $4\times 4$ matrices, which 
are a generalization of the $2\times 2$ matrices
appearing for skyrmions in a single layer at $\nu=1$\cite{Fertig}.
In general, the SITE is characterized by the existence of 
10 independent order parameters. In our case of a ferromagnetic GS at $\nu$=2,
the wave-function (\ref{trialwf}) involves broken spin and isospin symmetries 
as given by the order parameters
\begin{eqnarray}
\langle c^\dagger_{m\sigma \Lambda } c_{m\sigma \Lambda '}\rangle 
\neq 0, \hspace{0.5cm} 
\langle c^\dagger_{m\sigma \Lambda } c_{(m+1)-\sigma \Lambda '} \rangle \neq 0. 
\label{ordparam}
\end{eqnarray}
From these order parameters we obtain the SITE energy, $E$, and the third 
components of both the SITE spin, $S=\sum_{m \sigma \Lambda } 
\langle c^\dagger_ {m \sigma \Lambda} c_{m \sigma \Lambda }-1 \rangle 
\sigma /2$, and isospin, $I=\sum_{m \sigma }\langle c^\dagger_
{m \sigma R}c_{m \sigma L}+c^\dagger_{m \sigma L}c_{m \sigma R} \rangle$. 

{\it Results.} We have performed calculations in the whole regime of 
parameters giving a ferromagnetic GS at $\nu=2$. The first important result 
of our Hartree-Fock calculation is the map of charged excitations, 
in terms of $\Delta_z$ and $\Delta_{sas}$, shown in Fig. 1 for a representative 
case $d=l$. In the SPE region, the quasiparticle is not bounded to any 
texture and the extra charge is shared by the two layers in a trivial 
symmetric SPE. The energy of this quasiparticle is
$(\Delta_z-\Delta_{sas})/2$, its spin $S$=0.5, and its isospin 
$I$=1. On the other hand, in the SITE region, 
we always find that $\gamma _3 \simeq 1$ and $\gamma _4 \simeq 0$. 
The extra charge is mostly located in one layer (here chosen as the left layer)
while the charge located at the right layer is  
less than 0.05 electrons for all the cases in the SITE region.
The existence of two different types of charged excitations  
also comes out from the numerical diagonalization for 9 
electrons that we have performed as mentioned above. 

The transition between SITE and SPE is abrupt as 
shown in Fig \ref{fig2}. The curves start at the $\Delta _{z}$ values 
in which the GS becomes ferromagnetic. For low $\Delta _{z}$ the spin of the 
SITE is large. When $\Delta _{z}$ increases, the higher cost 
in Zeeman energy implies a reduction of the size and spin of the 
quasiparticle. For a given value of $\Delta _{z}$, this energy cost is 
so large that the system changes abruptly to a SPE symmetric state with the 
spin having the lowest possible value ($1/2$). A very interesting result 
is the distribution of the SITE spin in the two layers. 
$S_R =\sum_{m \sigma } \langle c^\dagger_
{m \sigma R} c_{m \sigma R }-1 \rangle \sigma /2$, 
the part of the SITE spin located at the right layer,
is rather small as shown in Fig. \ref{fig2}. 
This reflects that the spin of the SITE is 
mainly due to the {\it parent} texture in the left layer. 

The dependence of the SITE on tunneling is given in Fig. \ref{fig3}. 
$S$ is shown as a function of $\Delta _{sas}$. All curves are cuted off at 
the values of $\Delta_{sas}$ where the GS changes from ferromagnetic
to canted antiferromagnetic\cite{Zheng,DasSarma}. 
At $\Delta _{sas}=0$, the layers are decoupled and the spin texture is 
completely localized in the left layer while the right layer is inert.
The excitation is exactly equal to a skyrmion for the 
same $\Delta _z$ in an isolated layer at $\nu =1$ \cite{Fertig}.
When tunneling is switched on (i. e. $\Delta _{sas} \neq 0$) 
$S$ increases up to a value which depends on $d$.

In order to get some insight on the internal structure of the SITE, 
we compute the spin in-plane components on each layer $(S_x+iS_y)_\Lambda$.
As sketched in the inset of Fig. 1, in both layers 
there is a winding number unity around the origin. Moreover, the {\it shadow} 
texture is dephased in $\pi $ with respect to the {\it parent} texture. 
This agrees with the antiferromagnetic character of the effective 
interlayer coupling present in the quantum field theory description of 
the DL system \cite{DasSarma}.  

The spin texture is accompanied by the appearance and 
increase of an isospin $I$ given in Fig. \ref{fig4}. 
$I$ increases significantly with tunneling.
The dependence of $I$ on $\Delta_z$ shows, once again,  
an abrupt change at the transition between SITE and SPE regions. 

The physical picture coming out from all the above results is the following: 

For small Zeeman and tunneling couplings, the system develops
a single layer spin texture, as intralayer exchange is the most 
important interaction. When tunneling
comes into play, one could expect this single layer quasiparticle to
eventually turn into a more symmetric spin texture, with
the extra charge shared by the two layers. Our results show that this is 
not the case and the reason is drawn from the careful analysis
of the energies competing in this problem. 
Exchange interaction plays a mayor role in keeping the extra
charge mostly bound to one of the layers. Since the intralayer exchange is 
much stronger than the interlayer one, the system prefers primarily to 
form intralayer textures. However, a purely intralayer spin
texture does not take any profit of the available tunneling energy.
Our results show that, although the charge transfer from the left layer to 
the right one is very small, tunneling allows to move a great amount of 
charge from antisymmetric to symmetric states. A finite isospin appears 
and energy is gained preserving, at the same time, the quasi-intralayer 
character of the spin texture as preferred by exchange interaction. 
Moreover, the new local distribution of charge in the right layer, actually 
screens the {\it parent} texture in the left layer. The SITE becomes 
smoother and its spin increases, overcoming the spin corresponding to a 
purely single layer texture for the same Zeeman coupling.
As $\Delta _z$ increases, the energy cost for creating an intralayer 
spin texture becomes eventually larger than the one corresponding 
to the trivial SPE. Then, the SITE shrinks abruptly into a SPE. 

{\it Experimental consequences.}
We have found that a bilayer system at $\nu=2$ in the ferromagnetic regime 
can have charge excitations with a spin larger than 1/2. This spin further 
increases with tunneling due to the formation of a SITE. Moreover, there is 
an abrupt change of the spin at the transition between the SITE and SPE regions 
of Fig. \ref{fig1}.  
The quasiparticle spin could be measured with the techniques used 
in establishing the existence of skyrmions in single layer at 
$\nu$=1: NMR\cite{Barret}, activation energy\cite{Schmeller}, 
optical absorption\cite{Aifer}, specific heat\cite{Bayot}.
In a particular sample, the Zeeman contribution could be varied either by 
applying a parallel magnetic field \cite{Schmeller} or 
by changing the $g-$factor by means of external pressure\cite{Maude,Leadley}. 
In this way, the abrupt transition SITE-SPE  
could be detected. An alternative should be the study of a  
set of samples with different $\Delta _{sas}$ in order to analyze 
the increase of the SITE with tunneling.

In summary, we present a variational wave-function for the description of charged 
excitations of a DL at $\nu =2$ when the GS is ferromagnetic. 
In the map of charged excitations, there is a SITE region where,  
even for rather large tunneling, the extra 
charge is mostly located in one of the layers. 
The possibility of tunneling between layers 
provokes an increase of the spin texture compared with 
that of a skyrmion in a single layer at $\nu$=1 with the same $\Delta _z$. 
This enlargement is accompanied by the appearance of an isospin texture. 
The characteristics of the SITE show a strong dependence on tunneling. 
The spin of the quasiparticle presents an abrupt transition when going from 
the SITE region to the SPE one by increasing $\Delta _z$. These features 
should allow the experimental observation of these new quasiparticles. 

We are indebted to J.J. Palacios, H.A. Fertig, A.H. MacDonald
and S. DasSarma for helpful discussions.
Work supported in part by MEC of Spain under contract No. PB96-0085, by the 
Fundaci\'on Ram\'on Areces and by the CAM under contract No. 07N/0026/1998.

%%\widetext
%%\narrowtext

\begin{figure}
\caption{Map of charged excitations in terms of $\Delta_z$ and $\Delta _{sas}$ 
(in units of $e^2/\varepsilon l$) for a representative case $d=l$. 
The shaded region corresponds to a non ferromagnetic GS for $\nu =2$. 
SPE stands for single-particle excitations while SITE stands for the  
spin-isospin textured excitations schematically depicted at the inset. 
{\it Parent} and {\it shadow} spin 
textures are rotated in $\pi $ with respect to each other. } 
\label{fig1}
\end{figure}

\begin{figure}
\caption{ Spin $S$ of the charged excitation 
as a function of $\Delta _{z}$ (in units of $e^2/
\varepsilon l$) for  $d=l$ and different values of $\Delta_{sas}$. 
The part $S_R$ of $S$ due to electrons located 
at the right layer is also shown. The abrupt transition 
separates the region (to the left) corresponding to a SITE from  
that (to the right) corresponding to a SPE.}
\label{fig2}
\end{figure}

\begin{figure}
\caption{ SITE spin $S$ as a function of  
$\Delta _{sas}$ (in units 
of $e^2/\varepsilon l$) for different values of $\Delta_z$ and $d$.}
\label{fig3}
\end{figure}

\begin{figure}
\caption{ Isospin $I$ of the SITE as a function: a) of $\Delta _{sas}$ 
for $\Delta_z =0.02 $, and  b) of $\Delta _{z}$ for $\Delta_{sas}=0.1$. 
$\Delta_z$ and $\Delta_{sas}$ are in units of $e^2/\varepsilon l$.
The abrupt transition in part b) separates the region (to the left) 
corresponding to a SITE from that (to the right) corresponding to a SPE.}
\label{fig4}
\end{figure}

%%\widetext


\begin{references}
\bibitem{note1}For the case of a hole, the trial wave-function should mix 
GS occupied orbitals with angular momentum $m$, with GS empty states 
with angular momentum $m-1$.
\end{references}
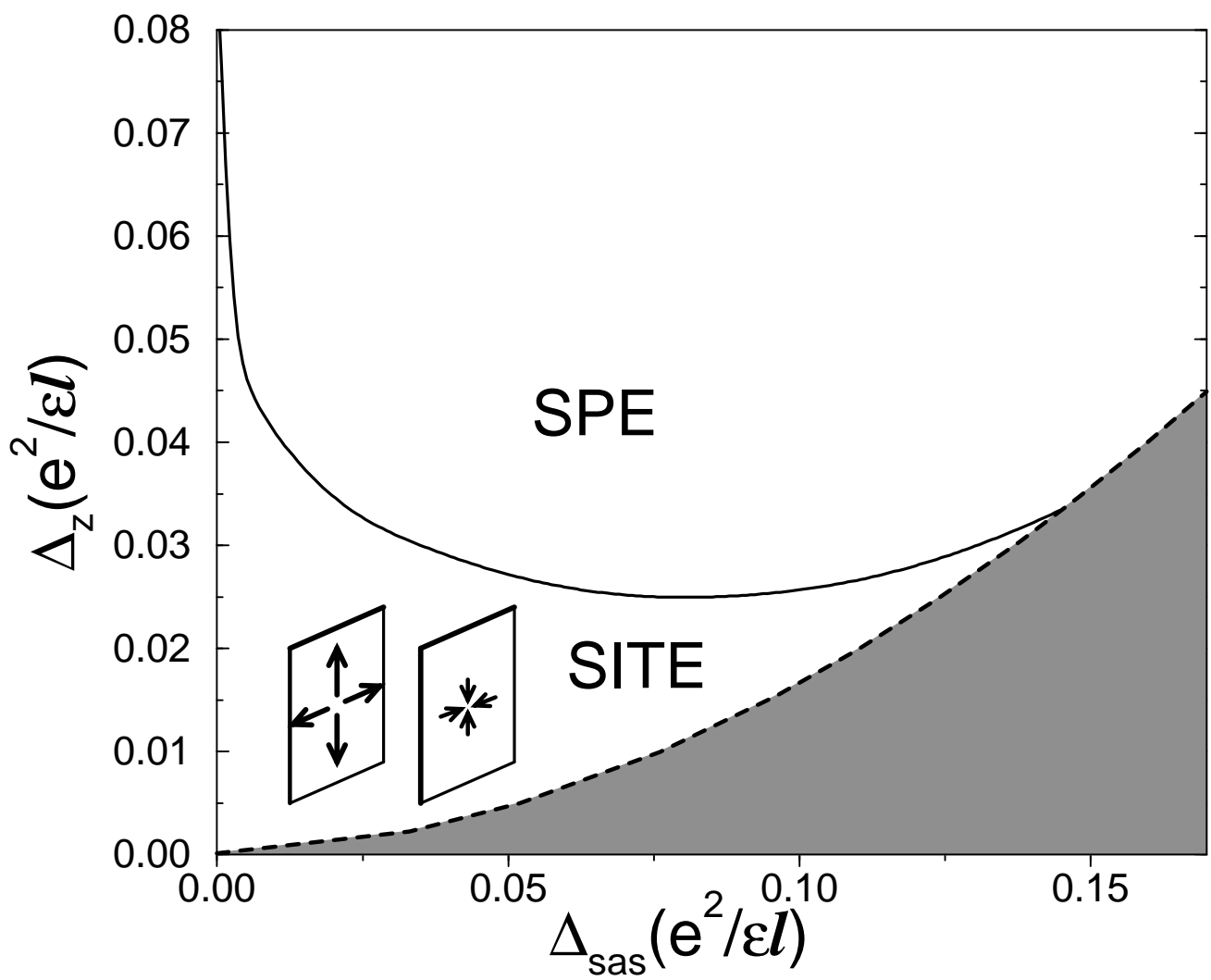
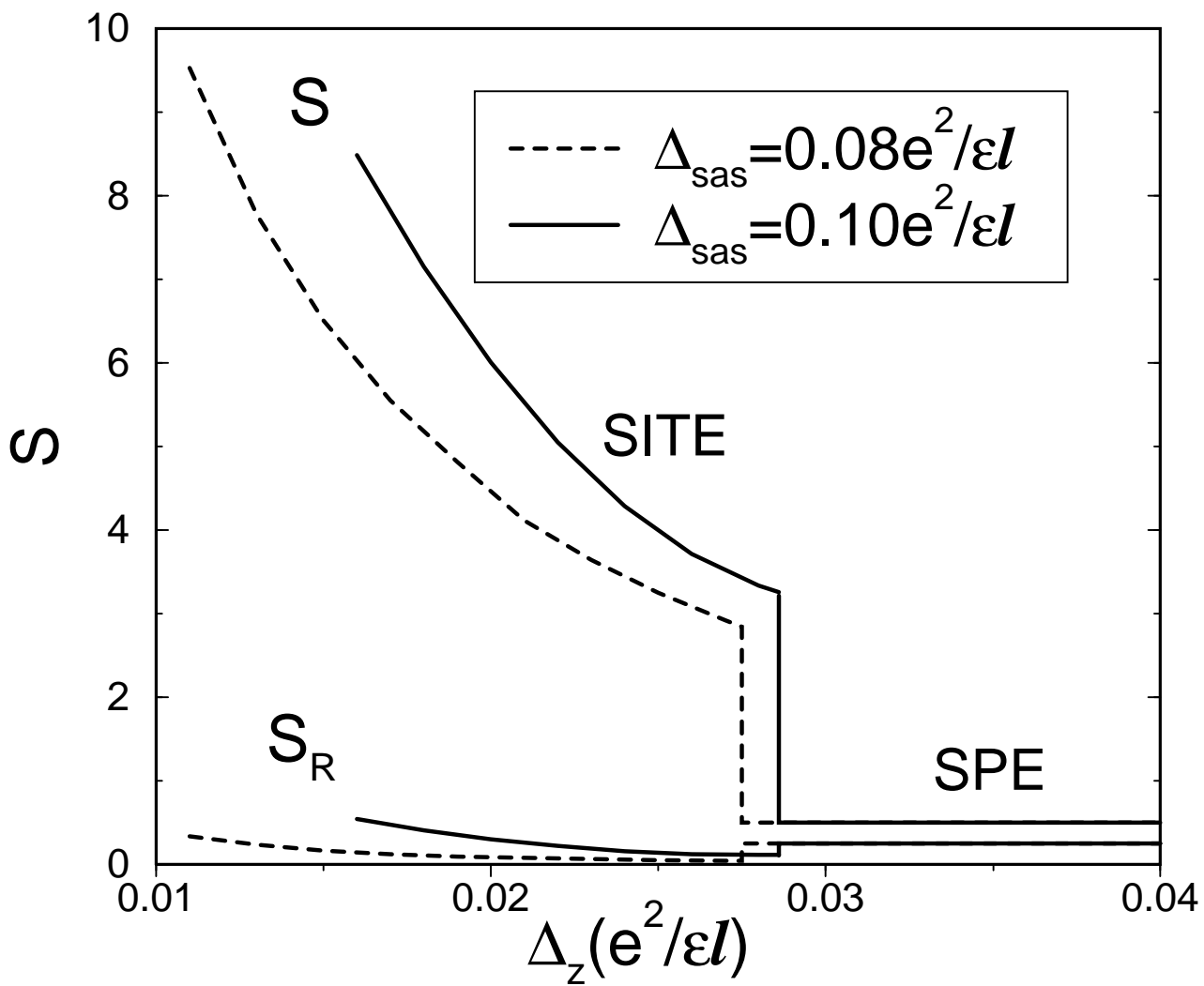
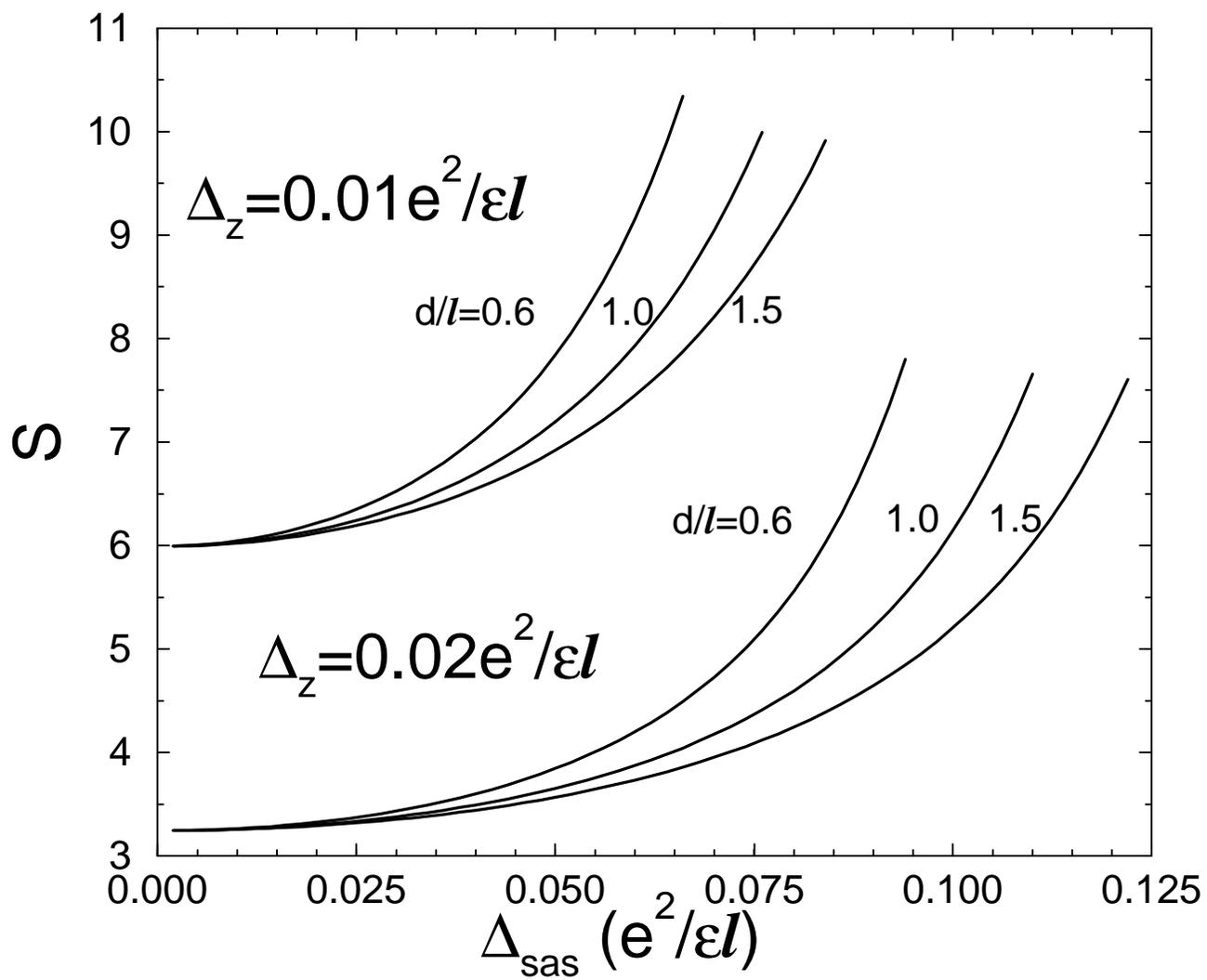
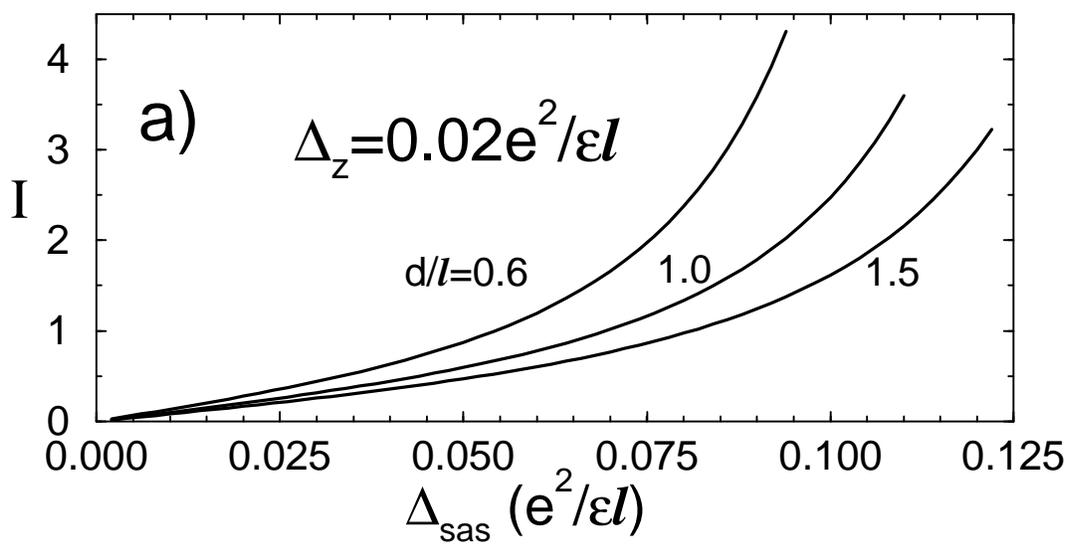
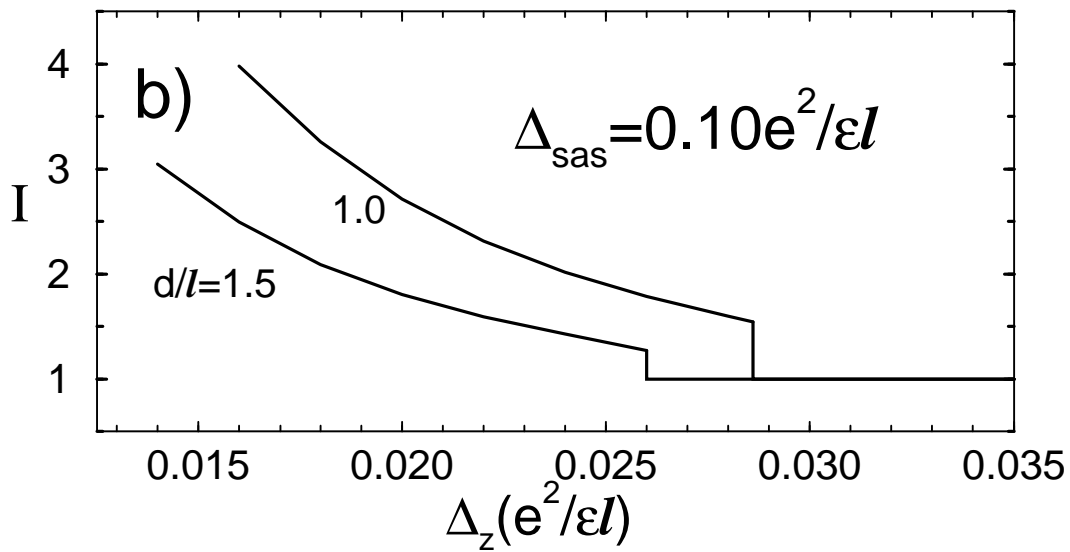
\end{document}